\documentclass{article}\usepackage{natbib,psfig,epsfig,emulateapj}
\newcommand{\beq}{\begin{equation}}
\newcommand{\eeq}{\end{equation}}
\newcommand{\bea}{\begin{eqnarray}}
\newcommand{\eea}{\end{eqnarray}}

\newcommand{\rem}[1]{ }
\bibpunct[,]{(}{)}{;}{a}{}{,}
\begin{document}

\title{Long-time evolution of magnetic fields in relativistic GRB shocks}

\author{Mikhail V. Medvedev{}$^{1,2}$,%\altaffilmark{*}, 
Massimiliano Fiore{}$^{3}$, Ricardo A. Fonseca{}$^{3}$, Luis O. Silva{}$^{3}$,
Warren B. Mori{}$^{4}$}
\affil{$^{1}$Department of Physics and Astronomy, 
University of Kansas, KS 66045}
%\altaffiltext{1}{Also at the 
\affil{$^2$ Institute for Nuclear Fusion, RRC ``Kurchatov
Institute'', Moscow 123182, Russia}
%\author{Massimiliano Fiore, Ricardo A. Fonseca, Luis O. Silva}
\affil{$^{3}$GoLP/Centro de Fisica de Plasmas, Instituto Superior T\'ecnico, 
 1049-001 Lisboa, Portugal}
%\author{Warren B. Mori}
\affil{$^{4}$Department of Physics and Astronomy, University of California, 
Los Angeles, CA 90095}

\begin{abstract}
We investigate the long-time evolution of magnetic fields generated
by the two-stream instability at ultra- and sub-relativistic astrophysical 
collisionless shocks. Based on 3D PIC simulation results, we introduce 
a 2D toy model of interacting current filaments. Within the framework
of this model, we demonstrate that the field correlation scale
in the region far downstream the shock grows nearly as the light crossing 
time, $\lambda(t)\sim ct$,  thus making the diffusive field dissipation 
inefficient. The obtained theoretical scaling is tested using numerical 
PIC simulations. This result extends our understanding of the structure of
collisionless shocks in gamma-ray bursts and other astrophysical objects.
\end{abstract}


%\maketitle

\section{Introduction}

Internal and external shocks in gamma-ray bursters (GRBs), internal 
shocks produced in jets of micro- and normal quasars and in active 
galactic nuclei jets, 
shocks in supernovae remnants, merger shocks in galaxy clusters and
large scale structure, --- all of them represent a single class of
strong collisionless shocks. The theoretical prediction that the 
counterstreaming instability operating at the shock produces 
strong magnetic fields \citep{ML99} has recently been confirmed 
in a number of state-of-the-art numerical sumulations 
\citep{Silva+03,Fred+04,Nish+04,Sakai+00,Bul+98}.
In this paper we investigate the long-time, nonlinear evolution
of the produced fields. The knowledge of the field dynamics 
downstream the shock is crucial for our understanding of the 
shock particle acceleration, as well as the spectral properties
(e.g., variability) of radiation produced at astrophysical 
shocks.

\section{The model}
\label{s:model}

Fully 3D PIC numerical simulations of the shocks demonstrate that the
generated magnetic fields are associated with a quasi-two-dimensional 
distribution of current filaments \citep{Silva+03}. Hence we suggest the
following toy model.

We consider straight one-dimensional current filaments oriented in the
vertical, $\hat z$, direction. Initially, all filaments are identical:
the initial diameter of them is $D_0$, their initial mass per unit length is
$\mu_0\simeq mn(\pi D_0^2/4)$, where $m$ is the mass of plasma particles 
(e.g., electrons) and $n$ is their number density. Each filament carries
current $I_0$ in either positive or negative $\hat z$-direction.
The net current in the system is set to zero, i.e., there are equal numbers
of positive and negative current filaments. We also assume that
the initial separation (the distance between the centers, see Fig. \ref{f:1}) 
of the filaments $d_0$ is comparable to their size, $d_0\simeq 2D_0$.
Finally, no external homogeneous magnetic field is present in the system.

\subsection{Single filament dynamics}

Initially, the filaments are at rest and their positions in space are random.
This configuration is unstable because opposite currents repel each other,
whereas like currents are attracted to each other and tend to 
coalesce and form larger current filaments.
The characteristic scale of the magnetic field will accordingly
increase with time. We study this process quantitatively using
the toy model of two interacting filaments. There are two limiting cases:
(i) when the characteristic velocity of the filaments during the coalescence 
process is much smaller than the speed of light and (ii) when these 
velocities  are comparable. We consider these cases separately.

\subsubsection{Nonrelativistic motion}

The magnetic field produced by a straight filament is $B_0(r)=2I_0/(cr)$, 
where $r$ is the cylindrical radius. The force per unit length acting on 
the second filament is $dF/dl=-B_0I_0/c$. Since $dF/dl=\mu \ddot x$, where
$x$ is the position in the center of mass frame and ``overdot'' denotes 
time derivative, we write the equation of motion as follows:
\beq
\ddot x=-\frac{2I_0^2}{c^2\mu_0}\,\frac{1}{x},
\label{eom}
\eeq
where we used that $r=2x$ and the reduced mass $\mu_r=\mu_0/2$.  
We define the coalescence time as the time required for the 
filaments, which are initially at rest, to cross the distance between 
them and ``touch'' each other, which happens when the distance
between their centers becomes equal to $D_0$, i.e., when $x=D_0/2$.
The coalescence time, as it is defined above, is independent of the details
of the merging process itself, which involves rather complicated 
dynamics associated with the redistribution of currents.
Quite obviously, the interaction between the filaments is the weakest
at large distances $x\sim x_0\sim d_0/2$. Hence, the coalescence rate 
is limited by the filament motions at the largest scales. 
The coalescence time can 
be readily estimated from Eq. (\ref{eom}), assuming that 
$x\sim x_0\sim d_0/2$ and $\ddot x\sim (d_0/2)\tau_0^{-2}$, as follows:
\beq
\tau_{0,NR}\sim\left({D_0^2 c^2 \mu_0}/({2 I_0^2})\right)^{1/2}.
\label{tau0NR}
\eeq 
The above estimate is valid as long as the motion is nonrelativistic. 
The maximum velocity of a filament is at the time of coalescence, $x=D_0/2$:
\beq
v_{m0}\sim D_0/2\tau_0\sim I_0/(c\sqrt{2\mu_0}). 
\label{vm0}
\eeq
It must always be much smaller than the speed of light.

It turns out that Eq. (\ref{eom}) can be solved exactly in terms
of time as a function of the filament separation:
\beq
t(x)={\sqrt{\pi\mu_0}\,cx_0}/({2I_0})\,
\textrm{erfc}\!\left(\sqrt{-\ln(x/x_0)}\right),
\label{soln}
\eeq 
where $x_0=x(0)$ and 
$\textrm{erfc}(y)=2\pi^{-1/2}\int_y^\infty e^{-\eta^2}d\eta$
is the complimentary error function. The coalescence time can
be calculated exactly:
\beq
\tau_{0,NR}={\sqrt{\pi\mu_0}\,cD_0}/({4I_0})\,
\textrm{erfc}\!\left(\sqrt{\ln 2}\right)
\approx 0.11\sqrt{\mu_0}cD_0/I_0,
\label{soln-tau0NR}
\eeq 
which is shorter than our estimate, Eq. (\ref{tau0NR}), 
by a factor of seven. The velocity as a function of position is
\beq
v(x)={2I_0}/({c\sqrt{\mu_0}})\,\sqrt{-\ln(x/x_0)}.
\label{soln-v}
\eeq
Consequently, the maximum velocity is
\beq
v_{m0}={2I_0}/({c\sqrt{\mu_0}})\,\sqrt{\ln 2}
\approx 1.67 I_0/(c\sqrt{\mu_0}).
\label{soln-vm0}
\eeq

\subsubsection{Relativistic motion}

If the motion of a filament during the merger becomes relativistic, 
the separation cannot decrease faster than as
$t(x)\simeq x/c$.
Therefore, the coalescence time will be
\beq
\tau_{0,R}\simeq (d_0/2)/c = D_0/c.
\label{tau0R}
\eeq

\subsection{Collective dynamics of filaments}

We now consider the filament coalescence as a hierarchical process.
Suppose that initially the system contains $N_0$ current filaments,
with an average separation $d_0\sim 2D_0$. Each of the filaments 
carries current $I_0$, its diameter is $D_0$ and its mass per unit
length is $\mu_0$. For simplicity, we asssume that filaments coalesce
pairwise. 

Having the original ``zeroth  generation'' of filaments merged 
(the process takes about $\tau_{0,NR}$ or $\tau_{0,R}$ to complete),
the system will now contain $N_0/2$ of ``first generation'' filaments.
Each of these filaments carries current $I_1=2I_0$, has mass per
unit length $\mu_1=2\mu_0$, and the separation between them 
is $d_1=\sqrt{2}d_0$ (because the two dimensional number density of filaments 
decreased by 2). Since $\mu\propto D^2$, the filament size also increases
as $D_1=\sqrt{2} D_0$. Remarkably, this new configuration is  
identical to the initial one, but with the re-scaled parameters. 
Hence, the coalescence process is self-similar. 
The produced first generation  filaments will be interacting with 
each other and merge again to yield the second generation.
The coalescence process will then continue in a self-similar way.
Note that the coalescence times at each stage are not necessarily the same.

At the $k$-th merger level, i.e., after $k$ pairwise mergers, 
the filament current, its mass per unit length and its size are
\beq
I_k=2^k I_0,\qquad  \mu_k=2^{k} \mu_0,\qquad  D_k=2^{k/2} D_0,
\eeq
whereas the
filament separation is $d_k\sim D_k/2$. The coalescence time at $k$-th 
level may be estimated in the way described in the prevous section. 
Using Eq. (\ref{tau0NR}) or (\ref{soln-tau0NR}) and Eq. (\ref{tau0R}), 
we obtain
\beq
\tau_{k,NR}= \tau_{0,NR},  \qquad
\tau_{k,R}= 2^{k/2}\tau_{0,R}. \label{tau-k}
\eeq
Since the coalescence time is independent of $k$ while the filaments 
are non-relativistic, whereas the distance between them increases, 
the typical velocities of the merging filaments grow with time and,
will approach $c$. From Eq. (\ref{soln-vm0}), we obtain
$v_{m,k}=2^{k/2}v_{m0}$.
The transition from the non-relativistic regime to the relativistic one
occurs after about $k_*$ mergers:
\beq
k_*=2\log_2(c/v_{m0}),
\label{k*}
\eeq
where $v_{m0}$ is set by the initial state of the system, 
Eq. (\ref{soln-vm0}).

Our primary interest is the evolution of the transverse correlation length 
of the magnetic field. With the filamets being randomly distributed in space,
the characteristic scale on which the magnetic field fluctuates is
about half the distance between the filaments, i.e., 
$\lambda_{B,k}\simeq d_k/2$. Thus, 
\beq
\lambda_{B,k}\simeq2^{k/2} D_0.
\eeq

Finally, it is instructive to present the evolution of the parameters
as a function of physical time, $t$, rather than the merger level, $k$.
Apparently, it takes $t=\sum_{k'=0}^k\tau_{k'} $
to complete $k$ mergers, where $\tau_k$ is given by Eq. (\ref{tau-k}). 
This equation implicitly defines $k(t)$.
In the non-relativistic case, the solution to this equation is
obvious: $k=t/\tau_0$, because $\tau_0$ is independent of $k$.
In the relativistic case, the sum of the series is easily calculated
using that $\sum_{i=0}^n a^i=(a^{n+1}-1)/(a-1)$. Thus, for the 
non-relativistic and relativistic cases respectively, the solution
to the equation $t=\sum_{k'=0}^k\tau_{k'} $ reads
\bea
k&=&{t}/{\tau_{0,NR}}, \label{kNR} \\
k&=&2\log_2\left[({t}/{\tau_{0,R}})(\sqrt{2}-1)+1\right]-1.
\label{kR}
\eea
Thus, the magnetic field correlation length increases as a function of
time as
\beq
\lambda_B(t)=D_0 2^{{t}/({2\tau_{0,NR}})}, \qquad
\lambda_B(t)\simeq  ct, \label{lambda}
\eeq
in the non-relativistic and relativistic regimes, respectively. Note that
the last expression is an approximation at large times $t\gg\tau_{0,R}$,
i.e., at large $k\gg1$.

\section{Application to collisionless shocks}
\label{s:shocks}

The growth rate and the saturation level of the field generated at shocks 
depend on the composition of the outflowing ionized gas 
\citep{ML99,Silva+03,Fred+04}. 
The gas composition in the cosmological outflows is not generally known.
Considering the interaction of the ejecta with the interstellar medium,
--- the external shock, --- it is quite reasonable to assume that the
shock is propagating through an electron-proton plasma. In contrast, the
ejecta itself, where internal shocks occur, may be either dominated by 
electron-positron pairs (leptonic jet) or by electrons and protons 
(baryonic jet) with or without $e^-e^+$-pair sub-population. 

Here, we will consider a general case of the ejecta containing
several species, labeled by the subscript $s$. We assume that the
species have different masses $m_s$, but their charges,
by the absolute value, are equal to $e$. In general, each species
has the bulk Lorentz factor (in the center of mass frame) $\gamma_s$
and the thermal Lorentz factor $\bar \gamma_s$, the latter represents the
random velocites of the particles (initially, $\bar \gamma_s< \gamma_s$). 
When the instability shuts off, 
the particles are randomized over the pitch angle, 
hence $\bar\gamma_s\simeq\gamma_s$.
In the context of astrophysical outflows, we assume that
all the species have the same $\gamma_s=\Gamma$ and $\bar\gamma_s\sim\Gamma$,
where $\Gamma$ is either the Lorentz factor of the external shock in 
observer's frame, or the {\em relative} Lorentz factor of the two colliding
shells measured in the 
frame comoving with the ejecta. In the latter case, the ejecta itself moves 
relativistically in observer's frame with the Lorentz factor 
$\Gamma_{\rm ej}$.

It is convenient to introduce the equipartition parameter for each species:
\beq
\epsilon_{B,s}=B^2_s/(8\pi n_s\gamma_s m_sc^2),
\eeq
 where $B_s$ is the 
strength of the magnetic field produced by the instability operating 
on the species $s$, $n_s$ is the number density of particles (both 
parameters are measured in the shock comoving frame). The equipartition 
parameter describes the efficiency of the magnetic field generation process, 
i.e., what fraction of the total kinetic energy of the particles of 
each species goes into the magnetic field energy. Note that this 
definition of the equipartition parameter is different from the 
commonly used definition, which describes the fraction of the 
total kinetic energy of the ejecta (summed over all species) that
goes into magnetic field. We define the plasma and Larmor frequencies as
\bea
\omega_{p,s}&=&\left(4\pi e^2 n_s/\gamma_s m_s\right)^{1/2},~\\ 
\omega_{B,s}&=&eB_s/\gamma_s m_s c.
\eea
For convenience, we also introduce their ``nonrelativistic'' counterparts
$\hat\omega_{p,s}\equiv\omega_{p,s}\gamma_s^{1/2}$ and
$\hat\omega_{B,s}\equiv\omega_{B,s}\gamma_s$.
It is useful to remember the following relation:
\beq
{\omega_{B,s}}/{\omega_{p,s}}=\sqrt{2\epsilon_{B,s}}.
\eeq

We may now represent the main results of Section \ref{s:model} 
in terms of plasma parameters. First, 
the initial separation between the filaments, $D_0$, must be comparable 
to the characteristic correlation length of the magnetic field produced
by the instability. This length at the onset of the instability is, 
in turn, set by the wavenumber of the fastest growing mode \citep{ML99}:
$\lambda_s\simeq \bar\gamma_s^{1/2}c/\hat\omega_{p,s}$, which is 
a factor of $(\bar\gamma_s/\gamma_s)<1$ smaller than the relativistic 
skin depth, $c/\omega_{p,s}$. However, we cannot set this scale $\lambda_s$
as $D_0$ because, as indicated by 3D PIC $e^-e^+$ simulations 
\citep{Silva+03}, $\epsilon_B$ is not constant in time at the beginning 
of the nonlinear filament interaction (at $t\sim10\omega_{p,s}^{-1}$). 
Hence the analysis of  Section \ref{s:model} is not applicable in such a
regime. In fact, it takes few plasma times, $\omega_{p,s}^{-1}$, 
in their simulations for $\epsilon_B$ to attain its asymptotic value.
The field correlation scale at this moment ($t\sim15\omega_{p,s}^{-1}$)
is somewhat larger than $\lambda_s$. We include this uncertainty 
via the parameter $\eta>1$ as follows:
\beq
D_0=\eta\, ({c}/{\omega_{p,s}})
=\eta\,({c\sqrt{\Gamma}}/{\hat\omega_{p,s}}).
\label{D0}
\eeq
Second, using $\oint{\bf B}\cdot d{\bf l}=(4\pi/c)I $, we express the
current $I_0$ in terms of the equipartition parameter, $\epsilon_{B,s}$, as
\beq
I_0=\eta\Gamma\sqrt{{\epsilon_{B,s}}/{2}}\,({m_s c^3}/{e}).
\eeq
Third, the mass per unit length of a filament must take into account
that the particle thermal motion is relativistic, hence the masses 
are $\bar\gamma_s m_s\sim \Gamma m_s$. We obtain
\beq
\mu_0=\Gamma m_s n_s \left(\pi D_0^2/4\right).
\eeq

The temporal evolution of the field correlation scale is determined
by Eq. (\ref{lambda}), where $\tau_{0,NR}$ 
is given by (\ref{soln-tau0NR}). The coalescence time may be written as
\beq
\tau_{0,NR}=\frac{\alpha_\tau\eta}{2}\,\omega_{B,s}^{-1}
=\frac{\alpha_\tau \eta}{\sqrt{8\epsilon_{B,s}}}
\,\omega_{p,s}^{-1}\sim\omega_{p,s}^{-1},
\label{plas-tau0NR}
\eeq
where we introduced the numerical factor 
$\alpha_\tau=(\sqrt{\pi}/4)\,\textrm{erfc}(\sqrt{\ln{2}})\approx0.11$.
Hereafter we assumed the typical values: $\epsilon_{B,s}\sim10^{-3}$
and $\eta\sim1$.The maximum merger velocity $v_{m0}$, Eq. (\ref{soln-vm0}), 
in terms of the speed of light is
\beq
v_{m0}/c=\alpha_v\sqrt{8\epsilon_{B,s}}\sim 0.1\, ,
\eeq
where another numerical factor is introduced: 
$\alpha_v=2\sqrt{\ln{2}}\approx1.67$.
The transition from the non-relativistic to relativistic coalescence 
regime occurs after $k_*$ mergers, given by Eq. (\ref{k*}). i.e., at the time 
\beq
t_*=2\log_2(c/v_{0m})\,\tau_{0,NR}\sim5\tau_{0,NR}.
\label{t*}
\eeq

\section{Numerical results}

We now compare our theoretical predictions with the results of particle-in-cell
numerical simulations. We used the numerical code OSIRIS,
described elsewhere \citep{Fonseca+02}, and we have performed 2D simulations 
($1280 \times 1280$ cells, $128.0 \times 128.0 \, (c/\omega_{p,e})^2$, 
9 particles/(cell$\times$species), 4 species) of the collision of 
weakly and fully relativistic neutral shells 
(electron-positron -- $e^- e^+$, and electron-proton -- $e^- p$) 
moving in the $\hat{z}$ direction, across the $xy$ simulation plane, 
with parameters similar to those in \cite{Silva+03}. 
%{\bf Can be taken out without problems: 
%For the relativistic shells in the simulation, the thermal speed is 
%actually smaller, which means that the filaments are more well 
%defined in the early stages of the simulation, thus making this 
%scenario closer to the theoretical model.}  
The temporal evolution of $\lambda_B$ as measured in the simulations 
is shown in Figure \ref{f:2}. Both a non-power-law nonrelativistic regime 
(until $t\approx 10-20/\omega_{p,e}$) and a power-law regime 
are clearly seen. The power-law fits
yield $\lambda_B(t)\propto t^\alpha$ with $\alpha\approx0.7$ for the
sub-relativistic cases ($u=0.6c$) and $\alpha\approx1.1$ for the 
relativistic cases ($\gamma=10$). Note also that the second power-law segment
with the same index is present at $t\ga100/\omega_{p,e}$ in 
$e^- p$ sub-relativistic run, indicating proton filament coalescences.
A similar behavior was also observed in 3D simulations 
(cf. \citealp{Silva+03}), but the significantly larger simulation 
planes presented here allow for improved statistics. 
At late times $t \gtrsim 100 /\omega_{p,e}$, the evolution of $\lambda_B$ 
rolls off and is slower: a return current is set-up around each filament, 
decreasing the range of the magnetic field to just a few electron 
collisionless skin depths (the thickness of the region where the 
return current is flowing), thus partially shielding the filaments. 
Filament coalescence then occurs at a slower rate.

\section{Discussion}

The two-stream instability generates magnetic fields at shock fronts 
very fast, with the typical $e$-folding time
 $\tau_{\rm growth}\sim\omega_{p,e}^{-1}\simeq10^{-5}(\Gamma/n)^{1/2}$~s,
where $n$ is the particle (e.g., electron) number density in cm$^{-3}$, 
which is of order of $10^{-10}$~s for internal GRB shocks and
is of order of $10^{-4}$~s for external shocks. The characteristic 
spatial scale above which the field is essentially random, as predicted 
by the linear instability theory, is
$\lambda_{\rm lin}\sim c\tau_{\rm growth}$, which is of order of
$10$~cm and $10^7$~cm for internal and external shocks, respectively.
The extremely short spatial 
scales, i.e., sharp field gradients, must be rapidly destroyed by 
dissipation. Indeed, it would be the natural result of pitch-angle 
diffusion of current-carrying charges in the chaotic magnetic fields. 
So, the question arises: Why the generated magnetic fields do not rapidly
decay back to zero as soon as the instability shuts off? The answer is:
The produced fields and the corresponding currents self-organize and form a 
quasi-two-dimensional distribution \citep{Silva+03}. A typical magnetic field
gradient scale grows with time very rapidly, 
with approximately the light crossing time $\propto t$; whereas the 
particle diffusion is a substantially slower process.
Hence, diffusive dissipation is drastically reduced. To illustrate this, 
we consider the field diffusion equation 
\beq
\partial_t B=-\kappa\partial^2_{xx}B
\eeq
with the dissipation coefficient, $\kappa$, being constant, for simplicity.
Approximating the spatial derivative as 
$\partial_x\sim\lambda_B(t)^{-1}\sim\lambda_0^{-1}(t/t_0)^{-\alpha}$,
where $\lambda_0$ and $t_0$ are constants and $\alpha>1/2$,
we obtain the scaling of $B$ with time as
\beq
B(t)=B_0\exp\left((t^{1-2\alpha}-t_0^{1-2\alpha})/\tau^{1-2\alpha} \right)
\to const.,
\eeq
as $t\to\infty$, where $\tau^{1-2\alpha}
=(2\alpha-1)\lambda_0^2/(\kappa t_0^{2\alpha})$.

We also note that in some respect, the field scale growth is analogous to the 
inverse cascade in two-dimensional magnetohydrodynamic (MHD) turbulence
(see, e.g., \citealp{B+B94}). The crucial difference is, however, the 
entirely {\em kinetic} nature of the process; at such small scales
$\sim c/\omega_p$ the MHD approximation is completely inapplicable.

To conclude, in this paper we analytically investigated the long-time
nonlinear dynamics of magnetic fields created at collisionless shocks
by the two-stream instability. We demonstrated that the field
correlation scale grows first exponentially and then nearly linearly with time,
see Eqs. (\ref{lambda}) and (\ref{plas-tau0NR}).
The transition from one regime to another occurs after few plasma times,
see Eq. (\ref{t*}). We compare our theoretical results with the
state-of-the-art PIC numerical simulation. Our fully 3D simulations 
\citep{Silva+03}
prove that the present simplified 2D analysis is accurate in the nonlinear 
regime until at least $t\sim {\rm few}\times 10^2/\omega_{p,e}$. 
Whether (or when) our 2D model breaks 
down at later times cannot be tested with the present computer capabilities.
The effect of the field evolution
on the observed spectrum and whether it can explain the spectral variability
of the prompt GRB emission deserves special consideration and will be
discussed in the subsequent paper.

\acknowledgements
One of the authors (MM) has been supported by NASA grant NNG-04GM41G 
and DoE grant DE-FG02-04ER54790. 
The work of some of the authors (MF, RAF, and LOS) is 
partially supported by FCT (Portugal).

\begin{figure}
\psfig{file=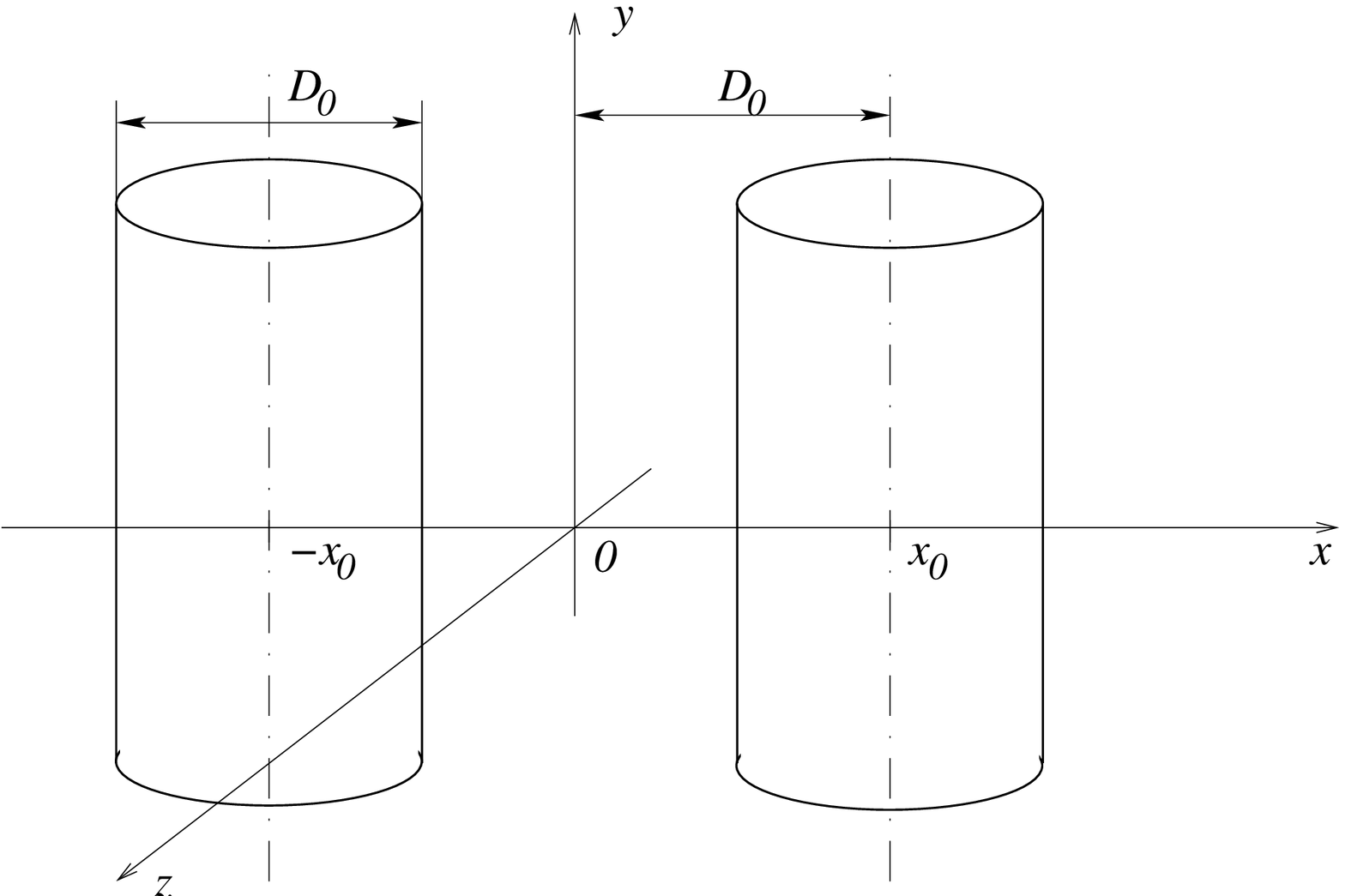,width=3in}
\epsfig{file=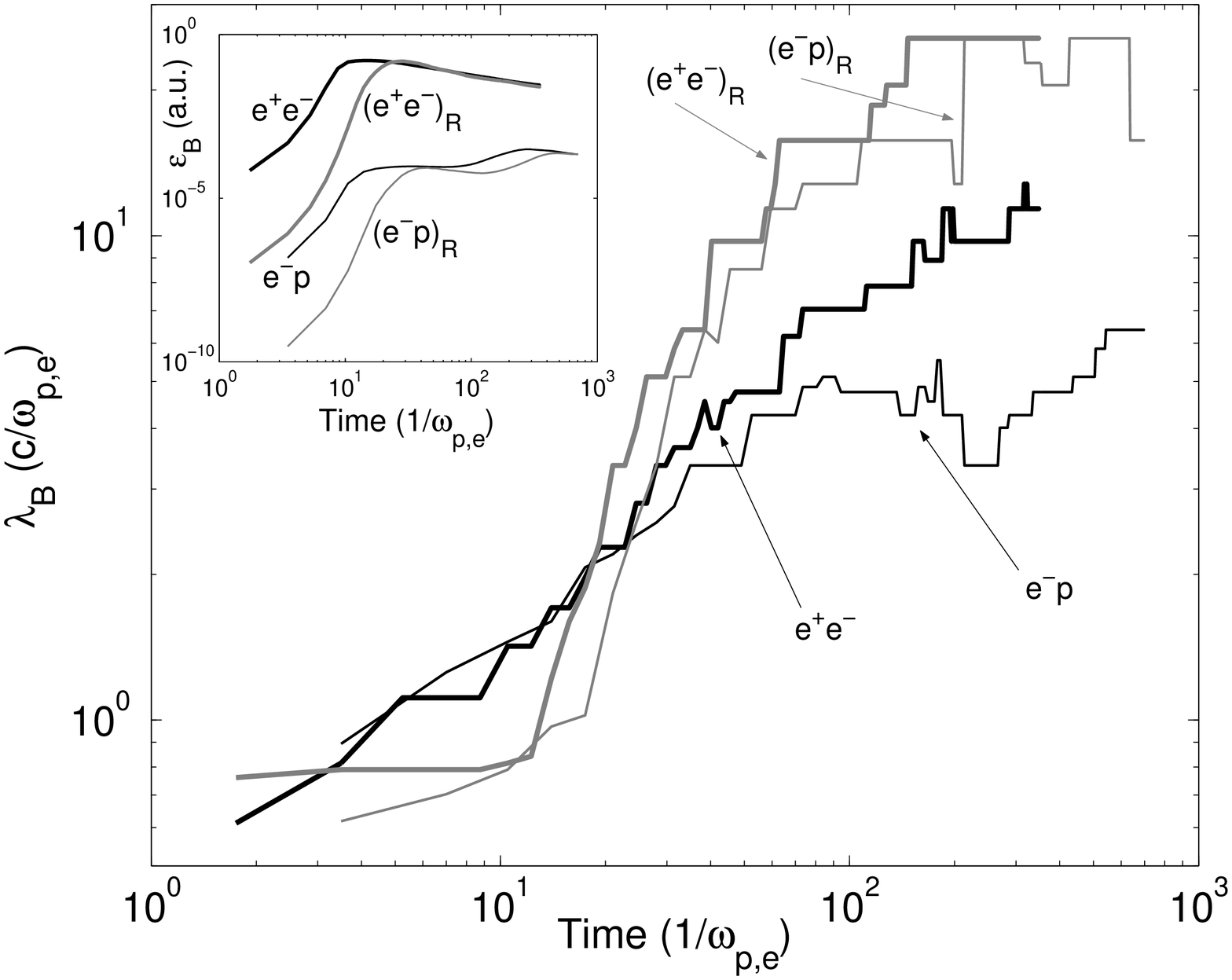,width=3in}
\caption{A schematic representation of the initial state.
\label{f:1} }
\caption{Temporal evolution of the field correlation length in four scenarios. 
The subscript R denotes relativistic shells 
($\mathbf{u}_R=\gamma \mathbf{v}=\pm 10.0 \, c\,\hat{z}$, 
while $\mathbf{u}=\pm 0.6 \, c\, \hat{z}$ otherwise, 
with $u_\mathrm{th} = \gamma v_\mathrm{th} =0.1 c$).
In the range $20/\omega_{p,e}<t< 100/\omega_{p,e}$, 
$\lambda_B \propto t^\alpha$ with 
$\alpha = 0.7,\, 0.6,\, 1,\, 1.2$ for 
$e^- e^+$, $e^- p$, $(e^- e^+)_R$, $(e^- p)_R$, respectively. 
In the inset, we show the evolution of the B-field equipartition parameter;
note that at $ 20/\omega_{p,e}\lesssim t \lesssim 100/\omega_{p,e}$, 
$\epsilon_B \approx \mathrm{const.}$.
\label{f:2} }
\end{figure}

\rem{
\begin{figure}
\psfig{file=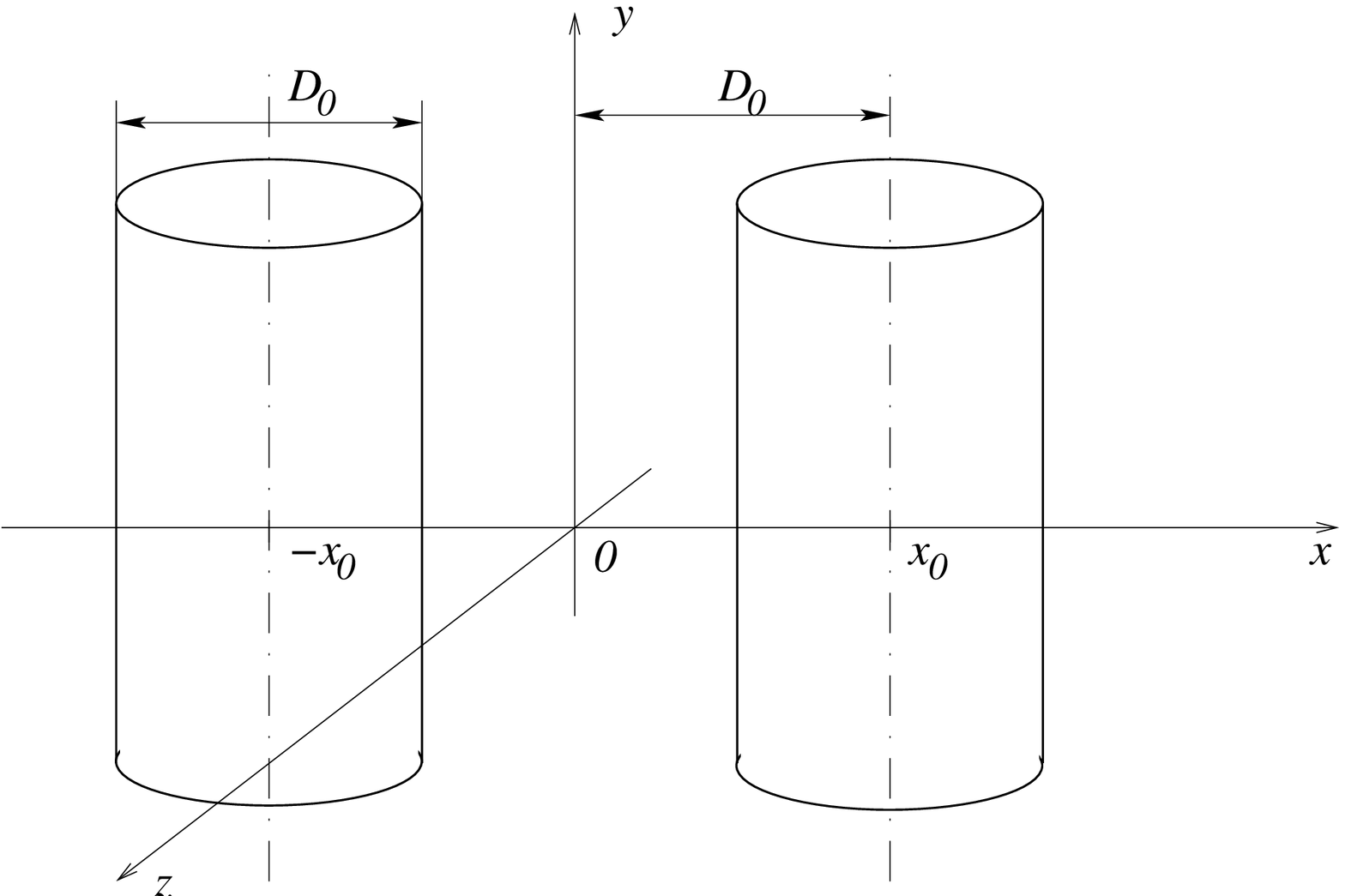,width=3in}
\caption{A schematic representation of the initial state.
\label{f:1} }
\end{figure} 
\begin{figure}
\epsfig{file=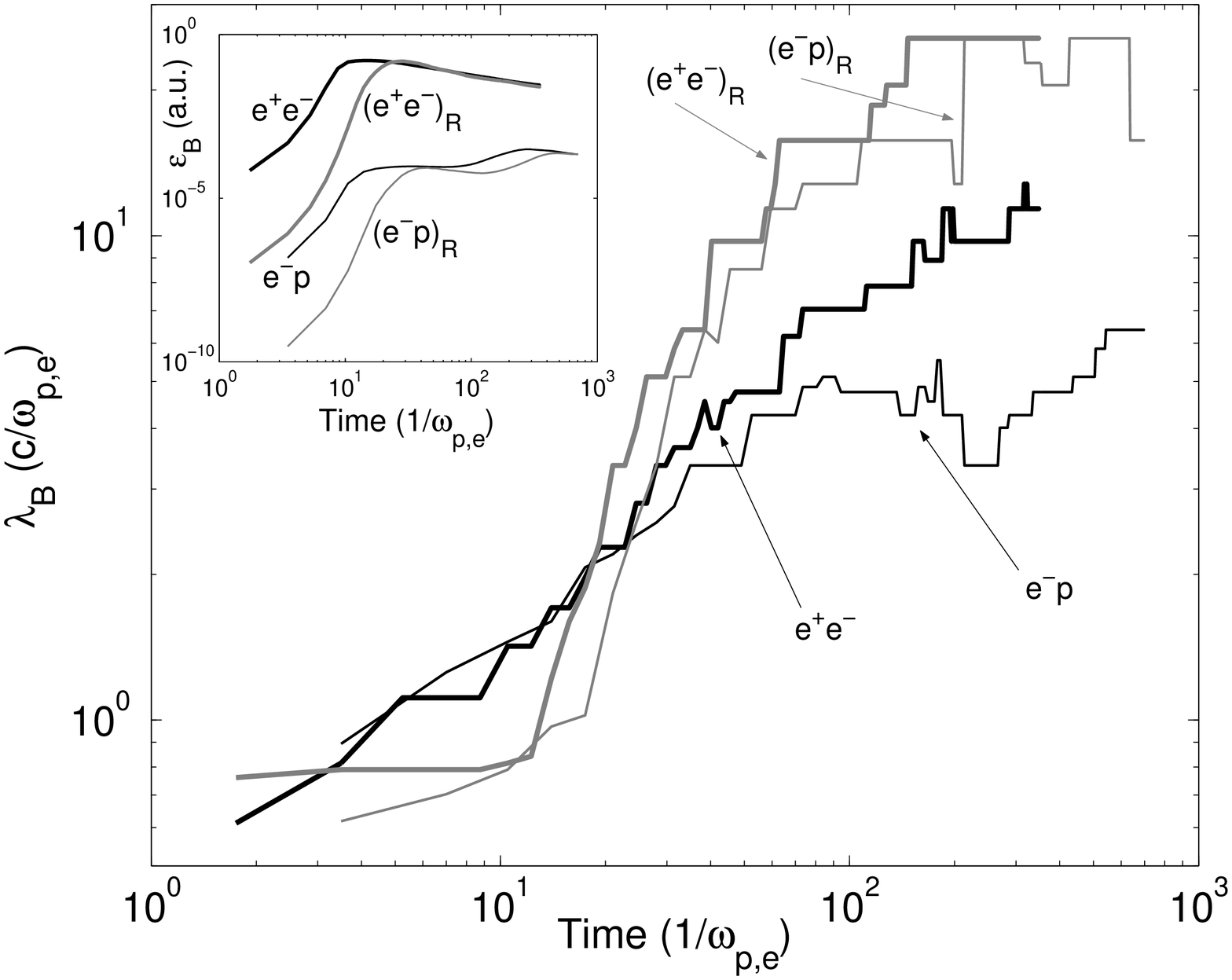,width=3in}
\caption{Temporal evolution of the field correlation length in four scenarios. 
The subscript R denotes relativistic shells 
($\mathbf{u}_R=\gamma \mathbf{v}=\pm 10.0 \, c\,\hat{z}$, 
while $\mathbf{u}=\pm 0.6 \, c\, \hat{z}$ otherwise, 
with $u_\mathrm{th} = \gamma v_\mathrm{th} =0.1 c$).
In the range $20/\omega_{p,e}<t< 100/\omega_{p,e}$, 
$\lambda_B \propto t^\alpha$ with 
$\alpha = 0.7,\, 0.6,\, 1,\, 1.2$ for 
$e^- e^+$, $e^- p$, $(e^- e^+)_R$, $(e^- p)_R$, respectively. 
In the inset, we show the evolution of the B-field equipartition parameter;
note that at $ 20/\omega_{p,e}\lesssim t \lesssim 100/\omega_{p,e}$, 
$\epsilon_B \approx \mathrm{const.}$.
\label{f:2} }
\end{figure} 
}

\end{document}